\documentclass[showpacs,preprintnumbers,amsmath,amssymb]{revtex4-2}
\usepackage{graphicx}% Include figure files
\usepackage{dcolumn}% Align table columns on decimal point
\usepackage{bm}% bold math
\usepackage{hyperref}% add hypertext capabilities
\usepackage{xcolor}

\begin{document}
    %\preprint{APS/123-QED}
    \title{The inflationary scenario in the $f(R)$ gravity model with a $R^4$ term}
    % Force line breaks with \\
    \author{Sahazada Aziz}
    \email{sahazada@outlook.com}
    \affiliation{Ramananda Centenary
        College, Laulara-723151, Purulia, West Bengal, India.}
    \author{Sohan Kumar Jha}
    \affiliation{Chandernagore College, Chandernagore, Hooghly, West
        Bengal, India}
    \author{Anisur Rahaman}
    \email{anisur.associates@aucaa.ac.in (Corresponding author)
        manisurn@gmail.com}\affiliation{Durgapur Government College, Durgapur - 713214,
        Burdwan, West Bengal, India}
    
    %\bibliographystyle{stylen}
    %\bibliography{ref.bbl}            %  but any date may be explicitly specified

    \begin{abstract}
        \begin{center}
            Abstract
        \end{center}
        We investigate the cosmic inflation scenario of a
        specific $f(R)$ model that contains more than one higher-order term in $R$. The
        $f(R)$ considered here has the terms $R^2$, $R^3$, and $R^4$ along with the linear term.
         A rigorous investigation has been carried
        out in the presence of these higher-order terms to figure out whether it
        leads to a physically sensible cosmic inflationary model. We examine in detail,
         subject to which conditions this $f(R)$ model renders a viable inflationary scenario,
         and it has been found that the outcomes of our study agree well with the recent PLANCK results.
    \end{abstract}

    \maketitle
\section{Introduction}
There has been a huge interest in the study of cosmic inflation
over the decades since it is considered as an effective scenario
for explaining  the origin of structure formation of the Universe.
There has been a huge interest in the study of cosmic inflation
over the decades since it is considered as an effective scenario
for explaining the origin of structure formation of the Universe.
In the article \cite{starobinsky1980new},
Starobinsky showed that Einstein's equations with quantum one-loop
contributions of conformally covariant matter fields admit a class
of nonsingular isotropic homogeneous solutions that correspond to
a picture of the Universe and it attracted a huge attentions since
successful slow-roll inflation can be achieved with a single
parameter which is the coefficient of the $R^2$ curvature term.
The predictions related to inflation of this model are well
consistent with the Planck data. Therefore, the discussion has
been generalized to a class of Starobinsky-like models having
common properties during inflation \cite{kofman1985inflationary,
kallosh2013universality, kallosh2014universal,
kallosh2013superconformal, kehagias2014remarks,
giudice2014starobinsky}. The Higgs inflation as a particular case
was studied in \cite{bezrukov2008standard}. In the article
\cite{kallosh2014universal}, it has been emphasized that there
must be a stage of inflation in  the early Universe to have a
consistent cosmological picture.The articles
\cite{muller1990power, gottlober1992models,ketov2011embedding}
contain the studies of different aspects of Starobinsky and
Starobinsky-like models. The Starobonsky model was extended with
the higher-order term in $R$ in the articles
\cite{sebastiani2014nearly, kamada2014topological,
artymowski2015inflationary}. Although the $\Lambda$$CDM$ model
provides a consistent explanation for the accelerating expansion
of the universe, the formation of large scale structure, cold dark
matter, and even for the most  mysterious dark energy, it still
suffers from the horizon, flatness, homogeneity,  and so-called
magnetic monopole problems. The elegant introduction of inflation
scenario eradicates these problems in a fascinating way
\cite{starobinsky1980new,  guth1981inflationary, linde1982new},
and the study of the cosmological observations on the cosmic
microwave background anisotropy have confirmed the predictions of
 inflation quite a long ago with a good degree of accuracy. The
most common framework for inflation is based on a scalar field
\cite{linde2005particle, liddle2000cosmological} that dominates
the energy density of the universe and rolls down slowly following
an almost flat potential $V(\phi)$. At the end of inflation, the
scalar field decays and the known environments for the standard
hot big-bang cosmology evolves out. This hypothetical scalar field
remains mysterious and to construct an  inflationary scenario, we
need either an extension of the standard model of particle physics
or alternatively, this scalar degree of freedom can be  thought of
as its origin lies in the gravitational sector. The theory that
includes a scalar field is the well known  Brans-Dicke theory. For
the vanishing Brans-Dicke parameter, this theory dynamically
corresponds to $f(R)$ gravity theory, where $f(R)$ is a general
function of the Ricci scalar \cite{sotiriou2010f, de2010f}. A
conformal transformation makes this theory equivalent to
Einstein's theory of  gravity which has minimal coupling to a
scalar field with a canonical kinetic term and a specific
potential described in terms of the function $f(R)$.  Whether this
scalar field, with a gravitational origin, can act as an inflation
 field that depends on the imposition of constraints on function
$f(R)$ such that it can offer a viable form of the potentials. In
this context, the slow-rolling conditions on $V(\phi)$ plays a
crucial role.

The Starobinsky inflation model \cite{starobinsky1980new}
attracted huge  attention in this respect because successful
slow-roll inflation can be  obtained with a single parameter
beyond the standard cosmological model,  namely the coefficient of
the $R^2$ term. The inflationary predictions of the Starobinsky
model are in excellent concordance with the recent  CMB anisotropy
data \cite{akrami2020planck,particle2020review,
martin2016observational,*chowdhury2019assessing} as well.
Therefore, extension, as well as  a generalization, been made to a
class of Starobinsky-like models with  common properties during
inflation
\cite{starobinsky1980new,kallosh2013universality,kallosh2014universal,
kallosh2013superconformal,kehagias2014remarks,
giudice2014starobinsky, muller1990power,gottlober1992models,
ketov2011embedding, sebastiani2014nearly,  kamada2014topological,
artymowski2015inflationary}, including Higgs inflation as a
particular case \cite{bezrukov2008standard, Park_2008}. Together
with this  framework, inflation can also, be generated in higher
dimensional theory through the compactification of extra
dimensions \cite{nakada2017inflation, ketov2017inflation,
otero2017r+}.

Extension of inflationary models with different $f(R)$ has been
reported after the pioneering work of Starobinsky
\cite{starobinsky1980new}. A logarithmic $f(R)$ has been used in
the article \cite{amin2016viable}. An inflationary model with
$f(r)$ having the form $R^2 + R^n$ is  discussed elaborately in
\cite{huang2014polynomial}. In the article \cite{kaneda2010slow},
$f(R)= R+ R^4$ is taken to construct an inflationary model. So the
studies of formulation of the inflationary model with different
choices of $f(R)$ and the necessary constraints on the free
parameter  used in the models have been enriching this field over
the years.  It should be mentioned that these higher order terms
of curvature  scalar $R$ start contributing if the spacetime
curvature is very high, which was expected to be in the very early
universe, i.e. in the vicinity of the Planck energy. The physics
of this high energy/curvature might be completely understood under
the framework of a string theory-like or supergravity-type UV
completing quantum gravity theory \cite{huang2014polynomial}. On
the  other hand, origin of these higher order terms can also be
realized  from the perspective of compactification of extra
spatial dimensions  of higher dimensional theories, where the
quartic term in $R$ is  naturally favored \cite{Chakraborty2016}.
In this sense the study  of inflationary scenario with $R^4$ term
is of worth investigation.

Starobnisky model was extended with $R^4$ term in
\cite{ketov2011embeddin} and it was shown that it led to
successful slow roll. In \cite{cheong2020beyond} an extension of
the Starobnisky model has been made with with $R^3$ term and it
has been established that it leads to a successful inflationary
model. Here we are intended deal with a generalized extension
considering the presence of both the $R^3$ and $R^4$ term. The
addition of higher order term although invites computational
difficulty however it would be of worth investigation whether with
the presence of both $R^3$ and $R^4$ is capable of leading to a
successful inflationary model. Whether the presence of $R^3$ term
along with $R^4$ spoil the slow roll is also a matter of
investigation.  In this article, an attempt, is therefore, made to
formulate an inflationary model with this specific $f(R)$ and
carry out a detail investigation towards constraining the free
parameters in order to make it a physically sensible framework for
cosmic inflation. We analyze inflation in both the original and
Einstein frames emphasizing that the scalar field picture is
essential to study the detailed consequences of inflation like the
number of e-folds, spectral index, tensor-to-scalar ratio and
reheating after inflation.

 The article is organized as follows. In Sec. II a general discussion
    of inflationary model based on $f(R)$ gravity is given. Sec.III. is
    devoted to the description of an inflationary model containing both
    $R^3$ and $R^4$ term with the evaluation of the potential. Sec.IV
    contains a discussion on constraining the free parameters to make
    an agreement with the available experimental data in order to make the model
    physically sensible with the evaluation of the slow-roll parameters,
    the number of e-folds, tensor to scalar ratio and the scalar spectral
    index. A discussion on the nature of potential is also presented in
    this section. In Sec.V, an estimation of the mass of the scalar field,
    the maximum reheating temperature and numbers of e-folds is made.
    In section VI, a  comparison of the results of the models constituted
    with three different $f(R)$ has been made to establish that the higher
    order term renders result with better precision  because of the freedom
    offered by extra free parameter. Finally, A brief summary of the work and
    a general discussion is given in Sec.VII.

    \section{Inflationary model based on $f(R)$ gravity}
 The general $f(R)$ theory of gravity, in a system of units where
    the reduced Planck mass ${M_{Pl}} \equiv {(8\pi G_N)^{ - 1/2}} =
    1$, is defined by the action
    \begin{equation}
        S=\frac{1}{2}\int d^4x \sqrt{-g} f(R) + \int d^4x {\cal L}(g_{\mu\nu}, \psi),  \label{ACT}
    \end{equation}
    where $f(R)= R + \tilde{f}(R)$.
   The determinant of the metric $g_{\mu\nu}$ is represented by $g$, $R_{\mu\nu}$ is the
   well known Ricci tensor and  $R= g_{\mu\nu}R^{\mu\nu}$.  The field $\psi_m$ represents
   the matter field. The field equation is obtained
   varying the action (\ref{ACT}):
   \begin{equation}
    f'R_{\mu\nu}-\frac{1}{2}g_{\mu\nu}+(g_{\mu\nu}\Box-\nabla_\mu\nabla_\nu)f'=T_{\mu\nu},
    \label{EQM}
 \end{equation}
where $f'=\frac{df}{dR}$ and the energy momentum tensor
$T_{\mu\nu}= \frac{1}{\sqrt{-g}}\frac{\partial {\cal L}}{\partial
g^{\mu\nu}}$. Trace of the field equation (\ref{EQM}) leads to
\begin{equation}
3\Box f'+ f'R-2f= T
\end{equation}
Where $T$ =  $g^{\mu\nu}T_{\mu\nu}$. This modified gravity
contains a scalar degrees of freedom which gets manifested when a
conformal transformation decouples it from metric to scalar field.
The conformal transformation leads to
\begin{equation}
S=\frac{1}{2}\int d^4 x\sqrt{g}[f(\phi) + f'(\phi)(R-\phi)
\end{equation}
Defining $\Phi= f'(\phi)$, and assuming the invertibility of the
relation this action can be written with a scalar
    field having minimal coupling with gravity as follows
    \begin{equation}
        S=\int d^4x \sqrt{-g}[\frac{1}{2}\phi R - U(\Phi)], \label{JA}
    \end{equation}
    where the potential is given by
    \begin{equation}
        U(\Phi)= \frac{(\Phi-1)\chi(\phi)-
            \tilde{f}\chi(\Phi)}{2},
    \end{equation}
    and frame function reads
    \begin{equation}
        \Phi \equiv 1 + \frac{\partial }{{\partial \chi }}[\tilde f(\chi)] .
    \end{equation}
    The action in the Einstein frame can be obtained by the
    conformal transformation $g^{\mu\nu}_{E}=\Phi g^{\mu\nu}$ which leads to
    \begin{equation}
        S=\int d^4x \sqrt{-g_E}[\frac{1}{2}\phi R_E
        -\frac{1}{2}g_{E}^{\mu\nu} -V(\phi)], \label{EA}
    \end{equation}
    where $V(\phi)$ has the expression
    \begin{equation}
        V(\phi)= \frac{1}{2\Phi^2}[(\Phi-1)\chi(\Phi)-
        \tilde{f}\chi(\phi)], \label{POT}
    \end{equation}
    and $\Phi$ and $\phi$ are related by the expression
    \begin{equation}
        \Phi= e^{\sqrt{\frac{2}{3}}\phi}. \label{PHI}
    \end{equation}
    Hereafter we will use Einstein frame for computation however we will
     omit the subscript 'E' for convenience. The slow-roll parameters
     $\epsilon$ and $\eta$ that follows from the action
    (\ref{EA}) respectively defined by \cite{liddle1992cobe,*liddle1994formalizing}
    \begin{equation}
        \epsilon  = \frac{1}{2}{\left( {\frac{{{V}^\prime }}{{{V}}}}
            \right)^2}, \label{EPSILON}
    \end{equation}
    \begin{equation}
        \eta  = \frac{{{V }^{\prime \prime }}}{{{V}}}.  \label{ETA}
    \end{equation}
    Here the symbol prime ($'$) denotes derivative with respect to
    $\phi$. In this framework the amplitude of primordial scalar power
    spectrum is defined by \cite{liddle1992cobe,*liddle1994formalizing,
    linde2005particle, liddle2000cosmological}
    \begin{equation}
        \Delta^2_R=\frac{1}{24\pi^2} \frac{V}{\epsilon}. \label{DEL}
    \end{equation}
    The number of times the universe expands during inflation to acquire the value $\phi_{end}$ starting from the value $\phi_N$ is
    known as the e-fold number which is known to have the expression \cite{liddle1992cobe,*liddle1994formalizing, linde2005particle, liddle2000cosmological}
    \begin{equation}
        N \equiv \ln \frac{{{a_{end}}}}{{{a_N}}} = \int_{{t_N}}^{{t_{end}}} H dt \simeq
        \int_{{\phi _{end}}}^{{\phi _N}} {\frac{V}{{V'}}} \,d\phi. \label{N}
    \end{equation}
    Here $a$ is the scale factor and $H$ represents the Hubble parameter. For this type of inflationary model there exist standard expression of scalar spectral index or scalar tilt $n_s$ and the tensor-to-scalar ratio $r$ with the slow-roll parameter $\epsilon$ and $\eta$ \cite{liddle1992cobe,*liddle1994formalizing}:
    \begin{equation}
        n_s= 1-6\epsilon+2\eta, \label{ns}
    \end{equation}
    \begin{equation}
        r= 16\epsilon.      \label{r}
    \end{equation}
    This is in short a of general discussion over $F(R)$ inflationary model. What follows next is the formulation of a new inflationary model with this input.
    \section{Inflationary model with a $f(R)$ containing both the $R^3$ and $R^4$ terms}
    In this article, an attempt has been made to formulate an inflationary model where $f(R)$ contains both the $R^3$ and $R^4$
    term along with the term $R^2$. The model with only $R^3$ term is studied in detail in \cite{cheong2020beyond}. Our work, in fact, is an
    extension over the model studied in \cite{cheong2020beyond} to investigate whether both the presence of $R^3$ and $R^4$ terms in the action
    can offer a successful inflation scenario. To this end, we consider
    \begin{equation}
        f(R)= R+\frac{R^2}{6M^2}+ \sum_{n=3,
            4}\frac{\lambda_n}{2n}\frac{R^n}{(3M^2)^n},
    \end{equation}
    where $M$ is the mass scale or energy scale of the theory and $\lambda$'s are the dimensionless coupling constants whose numerical values ${\lambda _{3,4}} \ll 1$, in order to evade the quantum gravity effects at the energy scale of inflation. It will be handy for computation if we express the above equation into the following
    \begin{equation}
        f(R)=
        R+\frac{\beta}{2}R^2+\frac{\gamma}{3}R^3+\frac{\delta}{4}R^4.
    \end{equation}
    The use of equation (\ref{PHI}) now leads to
    \begin{equation}
        f(R)= \beta\chi+\gamma\chi^2+\delta\chi^3 - (\Phi -1) = 0,
    \end{equation}
    which provides the following real solution for $\chi(\Phi)$:
    \begin{equation}
        \chi(\Phi)=\frac{Z}{3\sqrt[3]{2}\delta}
        -\frac{\sqrt[3]{2}(3\beta\delta-\gamma^2)}{3\delta
            Z}-\frac{\gamma}{3\delta},\label{CHI}
    \end{equation}
    where
    \begin{widetext}
        \begin{equation}
            Z=Z(\Phi, \beta, \gamma, \delta)= \sqrt[3]{\sqrt{(9\beta\gamma\delta-2\gamma^3+27\delta^2(\Phi-1))^2}
                +9\beta\gamma\delta-2\gamma^3+27\delta^2(\Phi-1)}.
        \end{equation}
    \end{widetext}
    We are now in a position to compute the potential $V(\phi)$ using the expression (\ref{POT},\ref{PHI}) with the input $\chi(\Phi)$ obtained in (\ref{CHI}). A lengthy but straightforward calculation leads to
    \begin{widetext}
        \begin{eqnarray}
            V(\phi)&&=\nonumber \\
            &&\left( {{e^{ - 2\sqrt {\frac{2}{3}}
                        \phi}}\left( { - 2\gamma + \frac{{2\sqrt[3]{2}\left( {{\gamma ^2} -
                                3\beta \delta } \right)}}{{\sqrt[3]{{\sqrt {{{\left( {9\beta
                                                    \gamma \delta
                                                    - 2{\gamma ^3} + 27{\delta ^2}\left( {{e^{\sqrt {\frac{2}{3}} \phi }} - 1} \right)} \right)}^2}
                                        - 4{{\left( {{\gamma ^2} - 3\beta \delta } \right)}^3}}  + 9\beta \gamma \delta  - 2{\gamma ^3}
                                    +27{\delta ^2}\left( {{e^{\sqrt {\frac{2}{3}} \phi }} - 1} \right)}}}} +}\right.} \right.\nonumber\\
            &&{\left. {{2^{2/3}}\sqrt[3]{{\sqrt {{{\left( {9\beta \gamma
                                            \delta - 2{\gamma ^3} + 27{\delta ^2}\left( {{e^{\sqrt
                                                        {\frac{2}{3}} \phi }} - 1} \right)} \right)}^2} - 4{{\left(
                                        {{\gamma ^2} - 3\beta \delta } \right)}^3}}  + 9\beta \gamma
                            \delta  - 2{\gamma ^3}
                            + 27{\delta ^2}\left( {{e^{\sqrt {\frac{2}{3}} \phi }} - 1} \right)}}} \right)^2}\nonumber\\
            &&\left( {6\beta  - \frac{{8{\gamma ^2}}}{{3\delta }}
                + \frac{{8\sqrt[3]{2}\gamma \left( {{\gamma ^2}
                            - 3\beta \delta } \right)}}{{3\delta \sqrt[3]{{\sqrt {{{\left( {9\beta \gamma \delta  - 2{\gamma ^3}
                                                + 27{\delta ^2}\left( {{e^{\sqrt {\frac{2}{3}} \phi }} - 1} \right)} \right)}^2} - 4{{\left( {{\gamma ^2}
                                                - 3\beta \delta } \right)}^3}}  + 9\beta \gamma \delta  - 2{\gamma ^3}
                                + 27{\delta ^2}\left( {{e^{\sqrt {\frac{2}{3}} \phi }} - 1} \right)}}}} + } \right.\nonumber\\
            &&\frac{{4\;{2^{2/3}}\gamma \sqrt[3]{{\sqrt {{{\left( {9\beta
                                            \gamma \delta - 2{\gamma ^3} + 27{\delta ^2}\left( {{e^{\sqrt
                                                        {\frac{2}{3}} \phi }} - 1} \right)} \right)}^2}
                                - 4{{\left( {{\gamma ^2} - 3\beta \delta } \right)}^3}}  + 9\beta \gamma \delta  - 2{\gamma ^3}
                            + 27{\delta ^2}\left( {{e^{\sqrt {\frac{2}{3}} \phi }} - 1} \right)}}}}{{3\delta }}\nonumber\\
            &+&\left( { - 2\gamma  + \frac{{2\sqrt[3]{2}\left( {{\gamma ^2}
                            - 3\beta \delta } \right)}}{{\sqrt[3]{{\sqrt {{{\left( {9\beta
                                                \gamma \delta
                                                - 2{\gamma ^3} + 27{\delta ^2}\left( {{e^{\sqrt {\frac{2}{3}} \phi }} - 1} \right)} \right)}^2}
                                    - 4{{\left( {{\gamma ^2} - 3\beta \delta } \right)}^3}}  + 9\beta \gamma \delta  - 2{\gamma ^3}
                                + 27{\delta ^2}\left( {{e^{\sqrt {\frac{2}{3}} \phi }} - 1} \right)}}}} + } \right.\nonumber\\
            &&{\left. {{2^{2/3}}\sqrt[3]{{\sqrt {{{\left( {9\beta \gamma
                                            \delta - 2{\gamma ^3} + 27{\delta ^2}\left( {{e^{\sqrt
                                                        {\frac{2}{3}} \phi }} - 1} \right)} \right)}^2} - 4{{\left(
                                        {{\gamma ^2} - 3\beta \delta } \right)}^3}}  + 9\beta \gamma
                            \delta  - 2{\gamma ^3} + 27{\delta ^2}\left( {{e^{\sqrt
                                        {\frac{2}{3}} \phi }} - 1} \right)}}} \right)^2} \nonumber\\
            &&/4\delta ))/864{\delta ^2}.  \label{V}
        \end{eqnarray}
    \end{widetext}
    This the exact expression of the potential that follows from the model with the generalized $f(R)$ considered here. Our next task is to constrain the free parameter space with the available experimental data to make this model physically sensible? The following section contains an account of that.
    \section{Constraints on the free parameters, calculations of slow-roll parameters  and predictions for $N$, $n_s$ and $r$}
\begin{figure}
        \includegraphics[width=\linewidth]{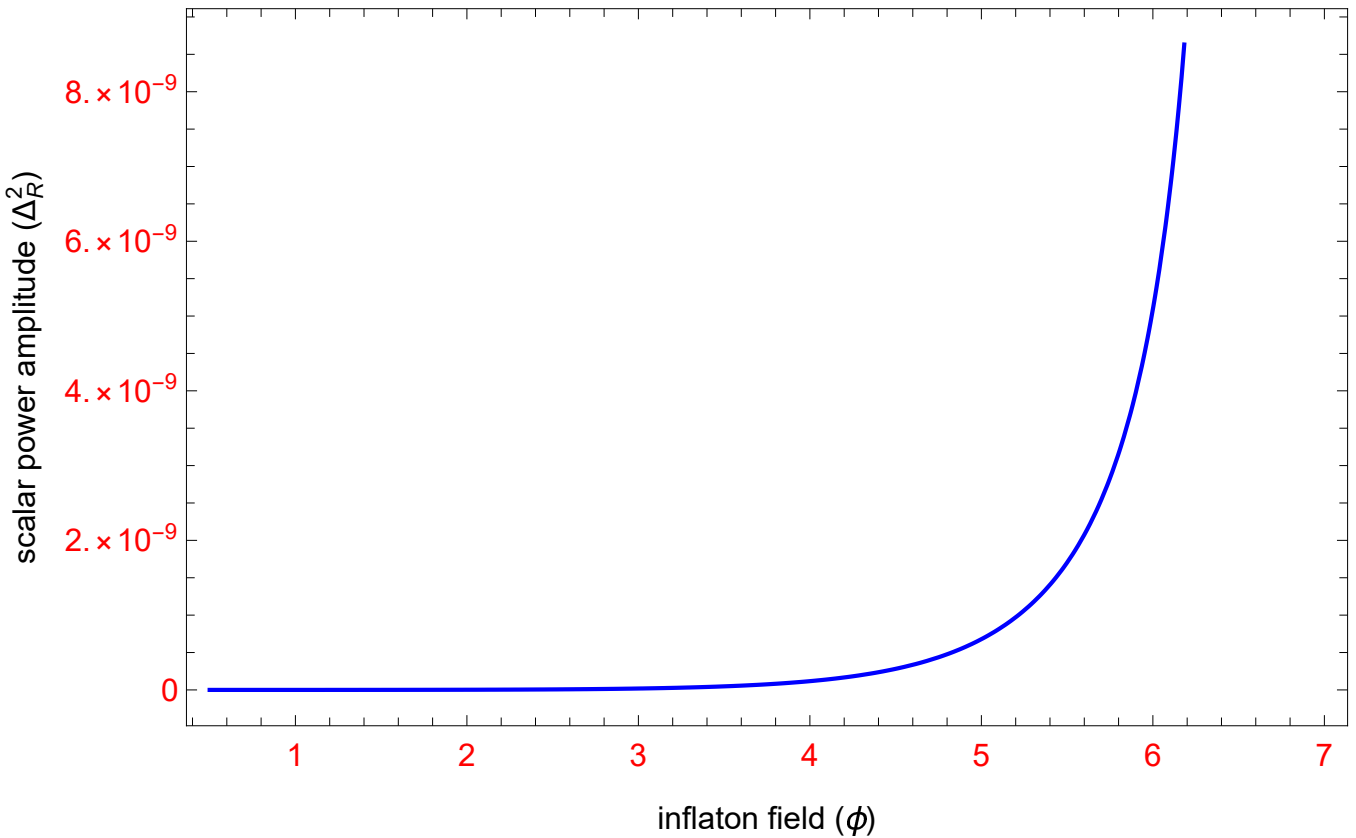}
        %\linewidth or \textwidth
        \caption{Plot of variation of scalar power amplitude ($\Delta_R^2$) with inflaton field ($\phi$)}.
        \label{fig:fig1}
    \end{figure}

\begin{figure}
        \includegraphics[width=\linewidth]{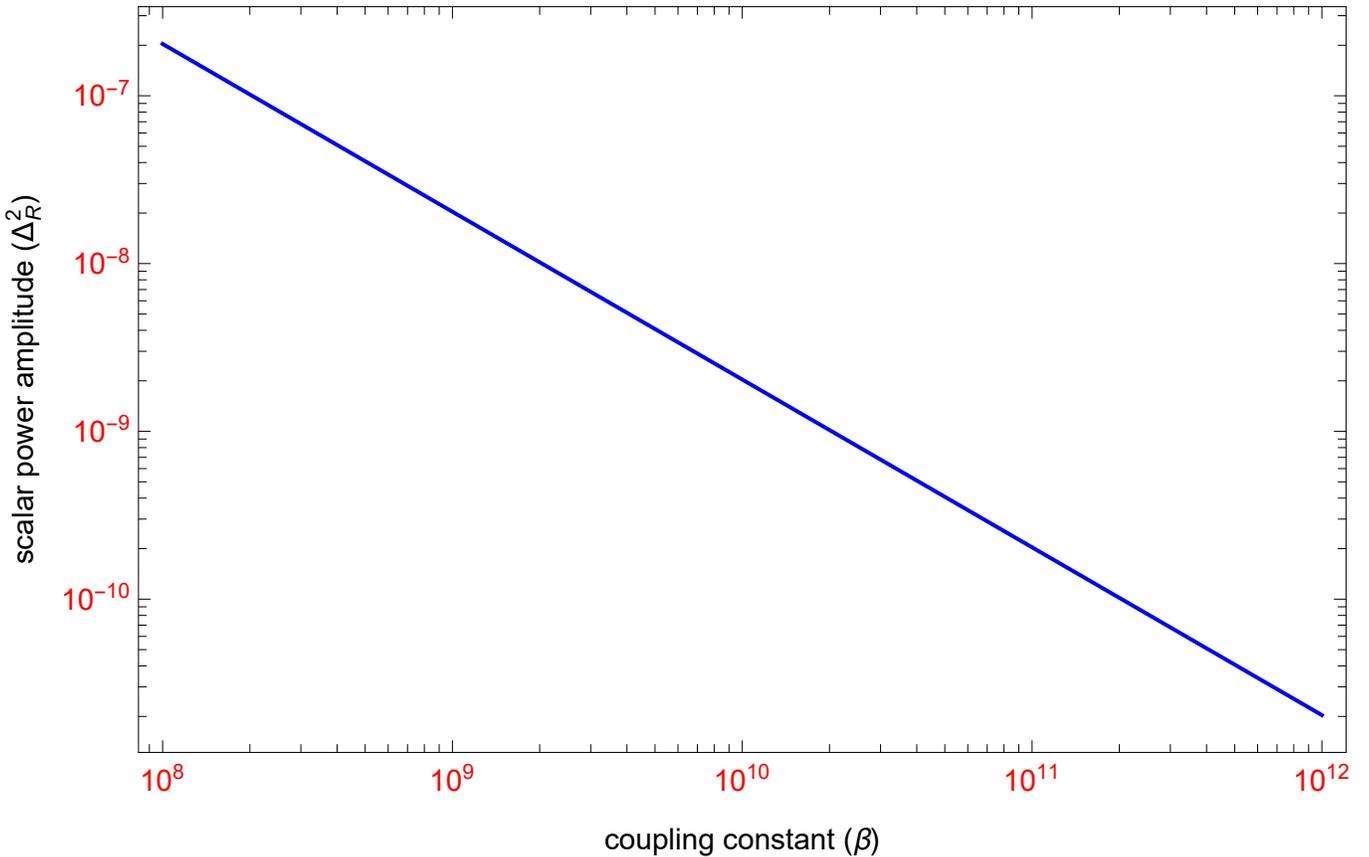}
        %\linewidth or \textwidth
        \caption{Plot of variation of scalar power amplitude($\Delta_R^2$) with coupling constant $\beta$}.
        \label{fig:fig2}
    \end{figure}

    The physical sensibility of an inflationary model   primarily depends on the fact whether it can exhibit   the prominent slow-roll of the inflationary scalar field called inflaton keeping the slow roll parameters within the admissible limit during inflation. In this respect let us recall the famous slow-roll conditions \cite{liddle1992cobe,*liddle1994formalizing}:
    \begin{eqnarray}
        \epsilon \ll 1, \nonumber \\
        |\eta|\ll 1.
    \end{eqnarray}
    The slow-roll parameters $\epsilon$ and $\eta$ are related to the
    measures of the slope and curvature of the potential respectively.
    The
    parameter  $\epsilon$ being a squared quantity always has a positive value. On the other hand, the sign of the parameter $\eta$ depends on the curvature of potential, so it can take both positive and negative signs. However, in order to compare the computed values obtained from a theory with the experimental result, we have to keep our focus at the time of \textit{horizon exit} or  \textit{Hubble crossing} \cite{liddle2000cosmological, liddle2003long,*martin2010first} where the potential  is almost flat to satisfy the slow-roll condition. Let us call
    the inflationary field at this point is $\phi_N$: value of inflaton at $N$ e-folds before the end of inflation. We will calculate the observable of our inflationary model at $\phi_N$, as the perturbations produced by inflation at this phase leave its signature on the observed CMB anisotropy \cite{akrami2020planck, particle2020review, martin2016observational, *chowdhury2019assessing, huang2014polynomial}.
    The theory with which we have started our investigation has three unknown free parameters $\beta$, $\gamma$, and $\delta$. In order
    to constrain the parameter space, we calculate the value of the amplitude of the primordial scalar power spectrum $\Delta_R^2$ by
    making use of Eq.\ref{DEL}. A study of the variation of $\Delta_R^2$ with the free parameter $\beta$, $\gamma$, $\delta$,
    and inflationary field $\phi$ will be useful in this respect. In Fig.\ref{fig:fig1} and Fig.\ref{fig:fig2} the necessary results have been furnished.
A careful observation reveals that the amplitude of the primordial
scalar power spectrum which is related to the COBE normalization,
    $\Delta^2_R = 2.1\times 10^{-9}$ \cite{akrami2020planck} corresponds
    to the parameter space of the theory with the constraints $\beta\simeq 10^{10}$ $\frac{\gamma}{\beta^2} \simeq 10^{-5}$, and $\frac{\delta}{\beta^3} \simeq 10^{-7}$ at $\phi_N \simeq
    5.5\,M_{Pl}$.
    We have mentioned that
    $\beta  \equiv \frac{1}{{3{M^2}}} = {10^{10}}$ corresponds
    to the energy scale or the mass scale of the inflationary model.
    The value of $\beta$ translates the mass scale of the theory to
    $M = 5.8 \times {10^{ -6}}{M_{Pl}} \simeq 1.4 \times {10^{13}}\,GeV$.
    We will use these particular values of the parameters throughout this paper.
    This is to be noted here that variation of $\beta$ from ${10^8}$ to ${10^{12}}$
    correspond to the variation of mass-scale
     $M$ from $5.7 \times {10^{ - 5}}\, M_{Pl}$ to $5.7 \times {10^{ -7}}\,M_{Pl}$.

\begin{figure}
        \includegraphics[width=\linewidth]{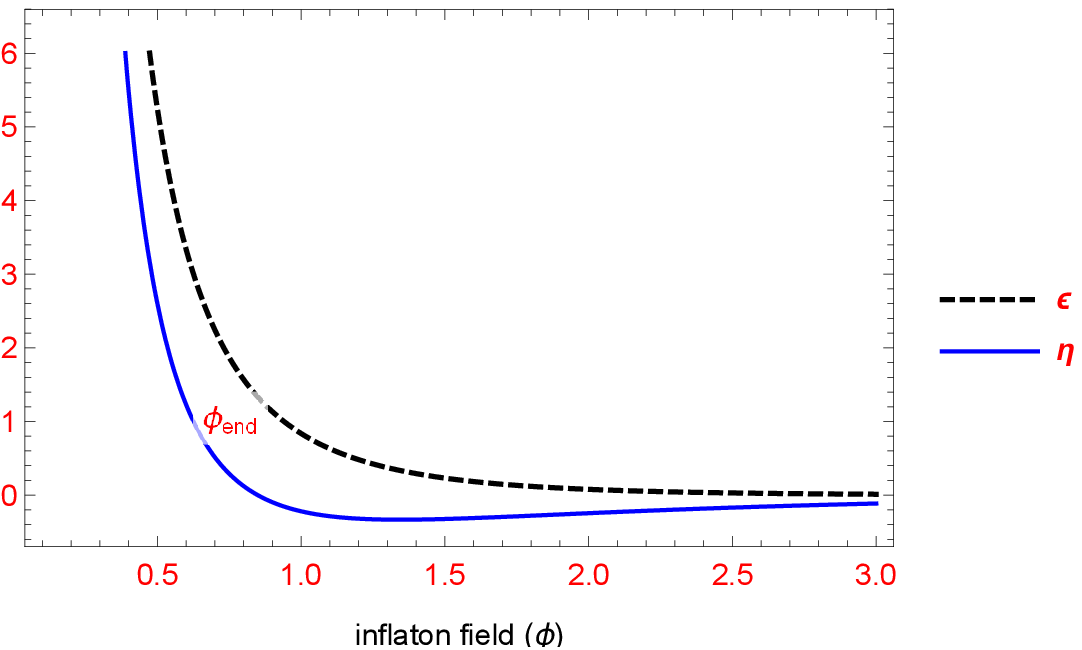}
        %\linewidth or \textwidth
        \caption{Plots of $\epsilon$ (black-dashed) and $\eta$ (blueline) with inflaton field ($\phi$) }
        \label{fig:fig3}
    \end{figure}

    The inflationary phase ends when either one or both the slow-roll conditions get violated. So $\epsilon = 1$ and/ or $|\eta|= 1$ can
    be safely considered as the end of inflation. Let us consider that the field at the end of inflation is $\phi_{end}$. Using the equations (\ref{EPSILON},\ref{ETA},\ref{V}) we compute  $\epsilon(\phi)$ and $|\eta(\phi)|$ varying  $\phi$ and this variation
    is graphically shown in the plots of Fig.\ref{fig:fig3}. From the plot  it is found that inflation ends at  ${\phi _{end}} \simeq 0.75\,{M_{Pl}}$.

    The information of termination point  ${\phi _{end}} \simeq 0.75\,{M_{Pl}}$ enables  us to calculate the number of e-folds using the equations (\ref{N},\ref{V}) with  the information of the starting point of inflation  $\phi_N =5.5 M_{Pl}$ which we have already in hand. The following integral
    \begin{equation}
        N = \int_{0.75}^{5.5} {\frac{V}{{V'}}} {\mkern 1mu} d\phi \simeq
        63,
    \end{equation}
    therefore, shows that the observable inflationary phase lasts for about 63 e-folds. We will examine in the next section whether this number is sufficient to solve the \textit{horizon problem}.

\begin{figure}
        \includegraphics[width=\linewidth]{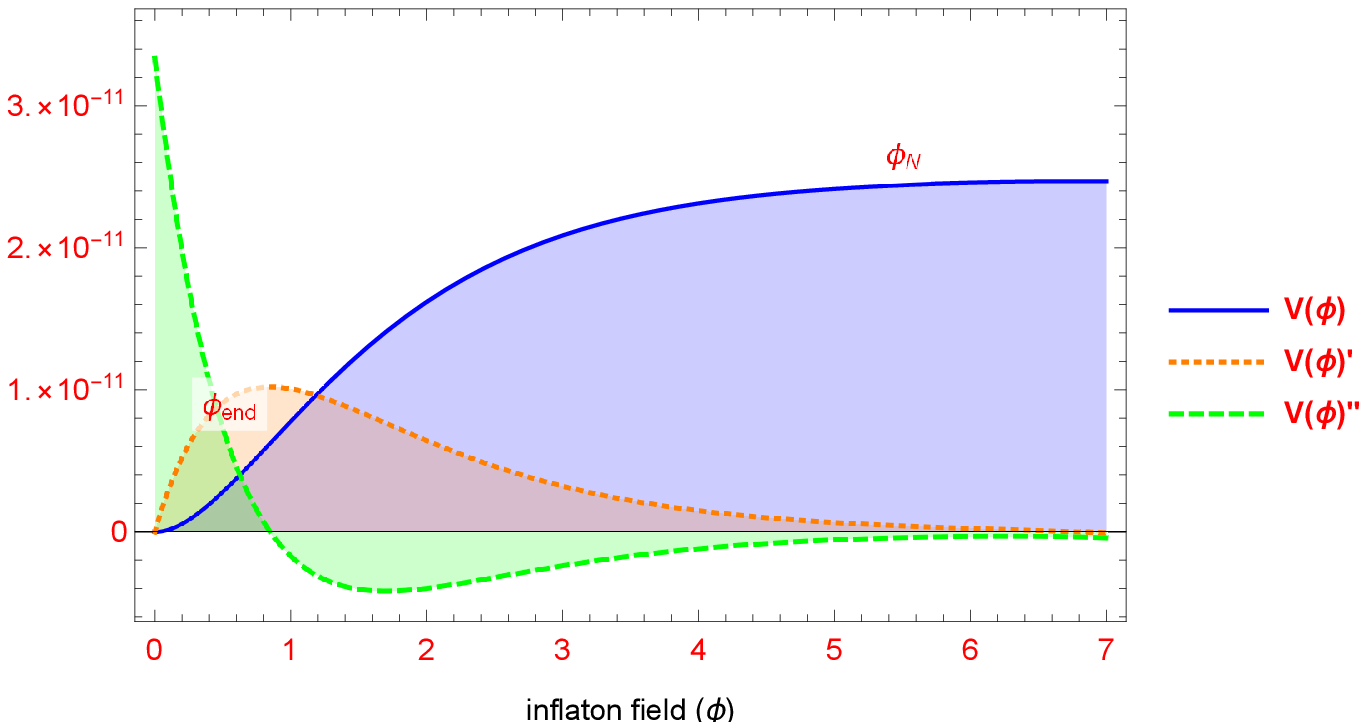}
        %\linewidth or \textwidth
        \caption{Plots of inflaton potential: $V(\phi)$ (Blue-solid), slope: $V(\phi)'$ (Orange-dotted) and curvature: $V(\phi)''$ (Green-dotted) with inflaton field $\phi$}.
        \label{fig:fig4}
    \end{figure}

    Now we are in a position to present a pictorial demonstration of the inflationary model that emerges out from the calculations we have carried out so far. It is shown in Fig.\ref{fig:fig4}.
    It shows that the \textit{observable inflation}, i.e. the phase of inflation of our interest, starts with $\phi_N =5.5 M_{Pl}$ and ends at around
    ${\phi _{end}} \simeq 0.75\,{M_{Pl}}$. It is worth mentioning that inflation initiates with a super-Planckian stage of the inflaton field $\phi$ and ends at the sub-Plackian stage. Total excursion of the field during inflation is $\Delta \phi \simeq 6.25\,{M_{Pl}}$ and during this phase the universe expands about 63 e-fold.

    We have given a plot of variation of the inflaton potential $V(\phi)$ with $\phi$.  The variation of the slope of the potential $V(\phi)'$ and the curvature of the potential $V (\phi )''$ with $\phi$ is also shown in the same plot. These three curves in Fig.\ref{fig:fig4}, plays an important role to estimate the value of the slow-roll parameters in one hand and helps to have an idea of how the experimental observable behave on the other. This nature of the inflaton potential mimics that of Starbonsky-Whitt potential \cite{linde2015inflationary}. The potential looks reasonably flat for the successful slow-rolling of the inflaton field. It is also to be noted that the slope of the curve is positive. If it is not so then the integral formula of Eq.\ref{N} fails. Inflationary models with such "Plateau type" potentials with concave curvature, $V''< 0$, are most preferred by cosmological observations \cite{akrami2020planck, martin2016observational, *chowdhury2019assessing} than the other types.

    Now, we calculate the values of scalar tilt $n_s$ and tensor to scalar ratio $r$ that our model yields. The values are respectively  given by:
    \[{n_s} \equiv {\left. {1 - 6\epsilon \left( {{\phi _N}} \right)
            + 2\eta \left( {{\phi _N}} \right)} \right|_{{\phi _N} = 5.5}}
    = {\rm{0}}.{\rm{966242}}\]
    and
    \[{\left. {r \equiv 16\epsilon \left( {{\phi _N}} \right)} \right|_{{\phi _N} = 5.5}} = 0.00242\]

\begin{figure}
        \includegraphics[width=\linewidth]{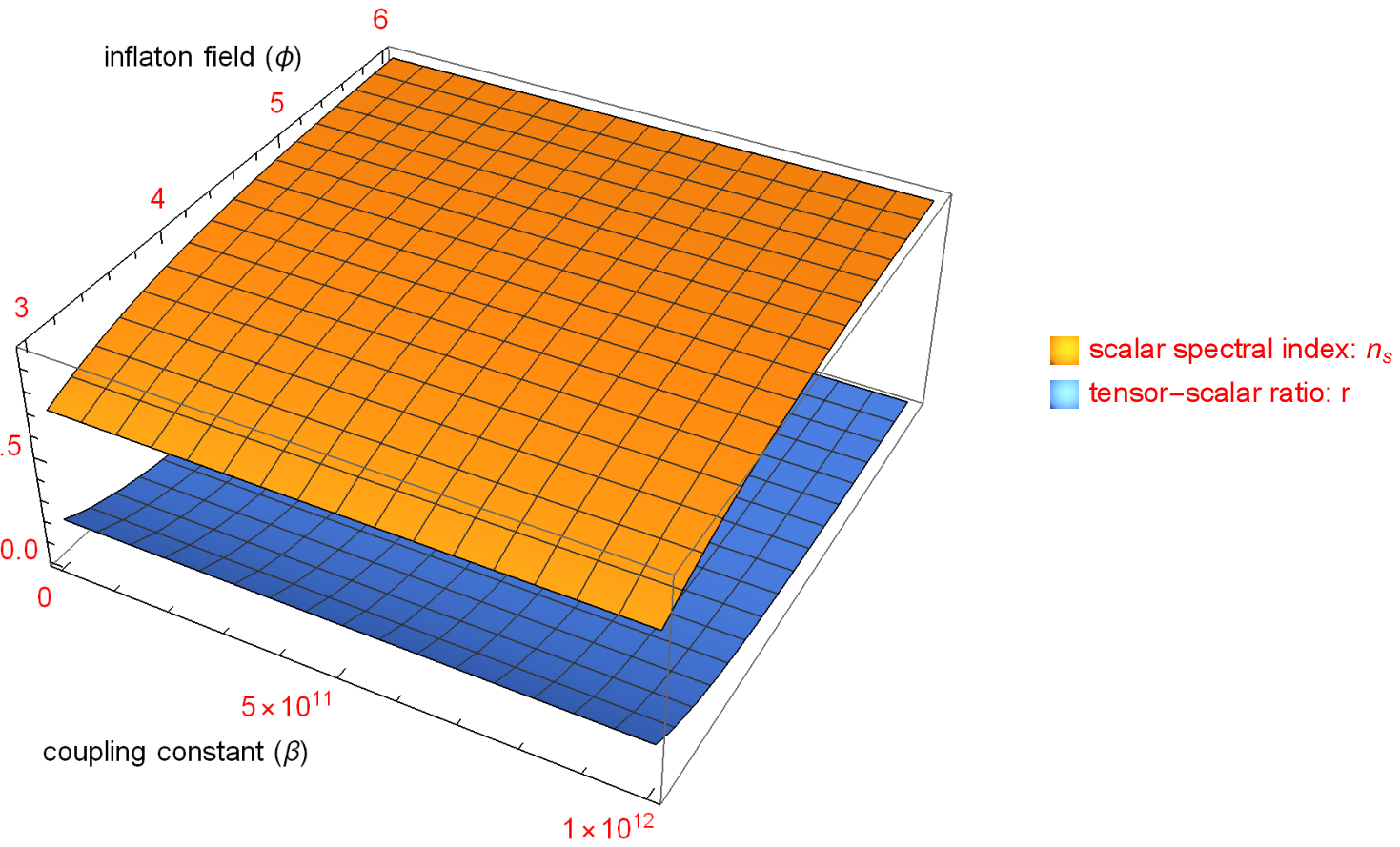}
        %\linewidth or \textwidth
        \caption{Plot of spectral index: $n_s$ (upper surface) and tensor to scalar ratio: $r$ (lower surface) with coupling constant $\beta$ and inflaton field $\phi$}.
        \label{fig:fig5}
    \end{figure}

    In Fig.\ref{fig:fig5} we exhibit the variation of $n_s$ and $r$ with respect to $\phi$ and $\beta$ respectively.
    From the figure, we can have an idea of how the spectral index and tensor-scalar ratio depends on $\phi$. The plots also exhibit the fact that the $n_s$ and $r$ are almost independent of the mass scale of the theory although the mass scale is fixed at $M = 5.8\times 10^{-6} M_{Pl}$ which is needed to obtain the primordial scalar power spectrum that agrees with the experiment. We observe that the value $r$ and $n_s$ are $0.00242$ and $0.966242$ respectively at $\phi_N = 5.5\,M_{Pl}$. These values are at par with the recent CMB observation \cite{akrami2020planck}, which tells that  ${n_s} = 0.9649 \pm 0.0042$ and $r < 0.06$. The predicted value of $r$ is also within Lyth bound \cite{LythPRL} which demands, approximately,
    \[r < 0.003{\left( {\frac{{50}}{N}} \right)^2}{\left( {\Delta \phi } \right)^2}.\] Putting $N=63$ and $\Delta \phi = 6.25$ we get the bound: $r < 0.073$. The value of $r$ predicted from our model is well within the Lyth bound.

    If we compare our results from the predictions inferred from classic formulae of Starobinsky model \cite{mukhanov1981quantum,*Starobinsky:1983zz}, it gives \[{n_s} \equiv 1 - \frac{2}{N} = 0.003023\] and \[r \equiv \frac{{12}}{{{N^2}}} = 0.968253\] for $N$=63. The small but noticeable difference in this result from the results we reported above is due to the inclusion of $R^3$ and $R^4$ terms.

    \section{Mass of the inflaton field, maximum reheating temperature and minimum number of e-folds}
    It would be instructive to get an estimation of the mass of the scalar field. For a viable theory, the mass of this scalar would be concomitant with the energy scale of the theory. Let us see how the mass of the scalar field can be estimated in this situation.
    \subsection{An estimation of the mass of the inflaton}
    It is reasonable to think without violating any physical principle that  the mass term of the inflaton field $\phi$ is present implicitly within the potential \ref{V}. If a potential has the following expansion
    \begin{equation}
        V(\phi ) = {c_1} + {c_2}{\phi ^2} + {c_3}{\phi ^2} + {c_4}{\phi
            _4} +  \cdot  \cdot  \cdot,
    \end{equation}
    the coefficient of the squared power of the $\phi$  gives the mass for the real scalar field $\phi$ which reads
    \begin{equation}
        m_\phi = \sqrt{2c_3}.
    \end{equation}
    at the classical level. Expanding the potential \ref{V} in a series of $\phi$ around $\phi =0$, we find that the mass of the field comes out to ${m_\phi} = 5.8 \times {10^{ - 6}}{M_{Pl}}\simeq 1.4 \times {10^{13}}\,GeV$ which is in agreement with the energy scale of the theory. This value of ${m_\phi}$ can be considered as a signature for our model. What happens when the inflation process terminates is known that the universe enters into the reheating phase. Let us have a glimpse of that with the estimation of maximum reheating temperature.

    \subsection{Estimation of maximum reheating temperature}
    When inflation comes to an end, the potential energy that causes the crucial slow-rolling starts to dissipate, setting the inflation field to oscillate quasi harmonically back and forth at the bottom region of the potential. As a result reheating  \cite{Kofman97,*motohashi2012reheating} of the universe gets started and that gives rise to the condition which becomes amenable to standard big bang cosmology with the creation of new particle . This state is known as the reheating phase. In this article, we will not study the reheating phase in finer detail but we will give an estimation of maximum reheating temperature of the Universe as suggested from this model.

    The inflaton scalar decays to all Standard Model(SM) particles. As the ${m_\phi } \gg {m_H}$, where ${m_H}$ is the Higgs mass, the dominant contribution of this decay comes from the SM electroweak sector \cite{Choi2019, cheong2020beyond}. The decay rate is given by \cite{Choi2019,*cheong2020beyond}
    \begin{equation}
        {\Gamma _\phi } \simeq \frac{{m_\phi ^3}}{{48\pi }} \label{Gamma}
    \end{equation}
    The reheating temperature, ${T_{re}}$ is related to this decay rate by the following relation \cite{cheong2020beyond}:
    \begin{equation}
        {T_{re}} \simeq {\left( {\frac{{90}}{{{\pi ^2}{g_*}}}} \right)^{1/4}}\sqrt {{\Gamma _\phi }} \label{Tre}
    \end{equation}
    Here $g_*$ is the number of relativistic degrees of freedom of the particles created at that time due to rapid oscillation after the termination of
    slow-rolling. Assuming all the SM particles relativistic at this energy scale, we take $g_*=106.5$. Substituting the value of ${m_\phi}$ in in \ref{Gamma} and using \ref{Gamma}, \ref{Tre} we obtain the value ${T_{re}} = 6.1 \times {10^{ - 10}}{m_{Pl}} \simeq 1.5 \times {10^9}\,GeV$.
    \subsection{Estimation of minimum number of e-folds}
    Having derived the value of maximum reheating temperature, ${T_{re}}$, we now evaluate the minimum number of e-folds required to solve the \textit{horizon problem}. The quantity is denoted by ${N_*}$ which is the number of e-folds from horizon exit to the end of inflation. A model independent calculation of ${N_*}$, in a matter dominated universe, leads to the expression \cite{Choi2016, cheong2020beyond,liddle2003long}:
    \begin{equation}
        {N_*} = 61.4 - \frac{1}{{12}}\ln \left( {\frac{{45{V_*}}}{{{\pi ^2}{g_*}T_{re}^4}}} \right) - \ln \left( {\frac{{V_{*}^{1/4}}}{{{H_*}}}} \right).       \label{N*}
    \end{equation}
    Here, ${H_*}$ is the Hubble rate that can be obtained from the relation\[{H_*} = {\left( {\frac{{{V_*}}}{3}} \right)^{1/3}}.\] In the above relation, ${V_*} \equiv {V_{end}} \equiv V({\phi _{end}}) = 5.24 \times {10^{ - 12}}$. Substituting all the values in \ref{N*} we finally obtain numerically the value of ${N_*} = 49.7 \simeq 50$. We obtained from an exact calculation, earlier in sec.IV, that the number of e-folds in this model gives $N=63$. So, It can be said that the value of $N$ obtained from this model is large enough to fit the requirement.

    \section{Brief comparison among $R+R^2$,
        $R+R^2+R^3$ and $R+R^2+R^3+R^4$ models}

    In this section, we discuss some of the salient features of thee $R+R^2$ and $R+R^2+R^3$ models and also provide a
    comparison of these two with the $R+R^2+R^3+R^4$ model in the same framework. For brevity, we use the notations $R2$, $R3$, and $R4$
    to denote $R+R^2$, $R+R^2+R^3$, and $R+R^2+R^3+R^4$ model respectively. The rest of the discussion in this section follows
    from the theoretical inputs described in Sec.III-V.

    In the original Starobinsky model $f(R)$ was taken as
    \begin{equation}
        f_{R2}(R)=R+\frac{R^2}{6M^2} =R+\frac{\beta}{2} R^2. \label{R2}
    \end{equation}
    With a $R^3$ correction to the above model one gets
    \begin{equation} f_{R3}(R)=
        R+\frac{R^2}{6M^2}+
        \frac{\lambda_3}{6}\frac{R^3}{(3M^2)^3}=R+\frac{\beta}{2}R^2+\frac{\gamma}{3}
        R^3.\label{R3}
    \end{equation}
    The real solutions $\chi(\Phi)$ (\ref{CHI})
    corresponding to the models (\ref{R2}) and (\ref{R3}) are
    \begin{equation} \chi _{R2}(\Phi ) = \frac{1 -
            \Phi}{\beta},
    \end{equation}
    \begin{equation} {\chi _{R3}}(\Phi ) =
        \frac{{\sqrt {{\beta ^2} + 4\gamma \,\Phi - 4\gamma } - \beta
        }}{{2\gamma }}
    \end{equation}
    respectively. We derive the inflaton potential for $R2$ model that reads
    \begin{equation}
        {V_{R3}}(\phi ){\text{ = }}\frac{{{e^{ - 2\sqrt {\frac{2}{3}} \phi
                }}{{\left( {{e^{\sqrt {\frac{2}{3}} \phi }} - 1}
                        \right)}^2}}}{{4\beta }}. \label{VR2}
    \end{equation}
    Similarly, we have the inflaton potential for$R3$ model:
    \begin{equation} {V_{R3}}(\phi ) =
        \frac{{{e^{ - 2\sqrt {\frac{2}{3}} \phi }}{{\left( {\beta - \sqrt
                            {{\beta ^2} + 4\gamma \left( {{e^{\sqrt {\frac{2}{3}} \phi }} - 1}
                                \right)} } \right)}^2}\left( {2\sqrt {{\beta ^2} + 4\gamma \left(
                        {{e^{\sqrt {\frac{2}{3}} \phi }} - 1} \right)} + \beta }
                \right)}}{{48{\gamma ^2}}}. \label{VR3}
    \end{equation}
    The known COBE normalization fixes the values of the coupling
    constants as $\beta\simeq {10^{10}}$ and $\gamma \simeq\frac{{{\beta ^2}}}{{{{10}^5}}}$.
    \begin{figure}
        \includegraphics[width=\linewidth]{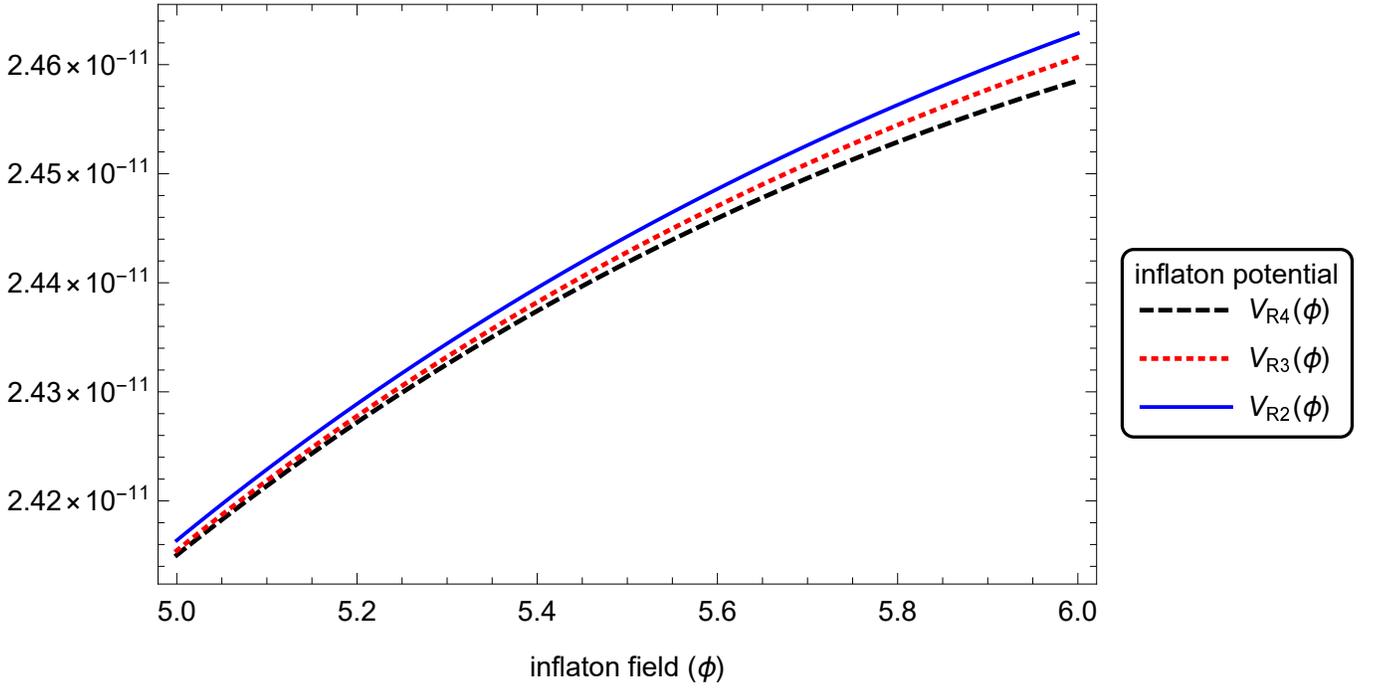}
        \caption{Plots of inflaton potential $V(\phi)$, for the models: $R4$ (Black-dashed), $R3$ (Red-dotted) and $R2$ (Blue-solid) with inflaton field $\phi$.}
        \label{fig:fig6}
    \end{figure}
    \begin{figure}
        \includegraphics[width=\linewidth]{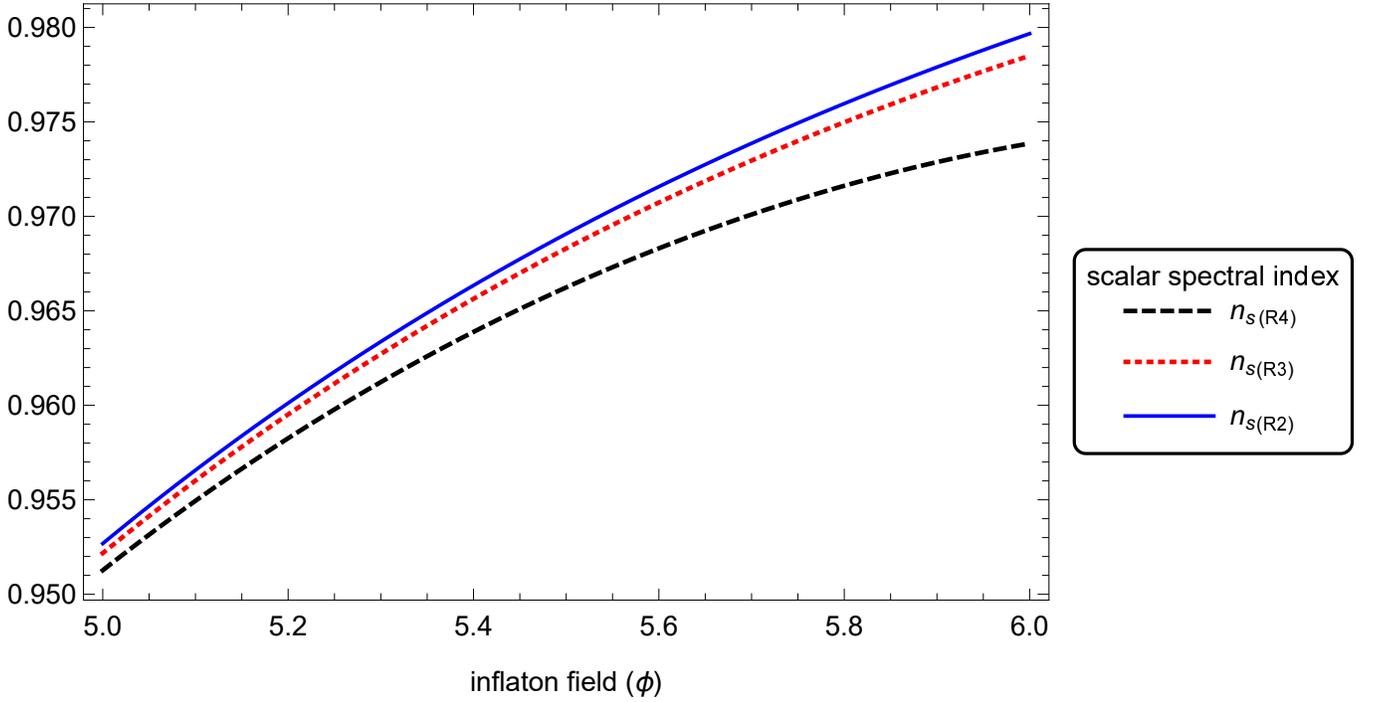}
        \caption{Plots of scalar spectral index $n_s$, for the models: $R4$ (Black-dashed), $R3$ (Red-dotted) and $R2$ (Blue-solid) with
            inflaton field $\phi$}.
        \label{fig:fig7}
    \end{figure}

    Now, we are in a position to compare the potentials obtained in Eqs.(\ref{VR2},\ref{VR3},\ref{V}). The plots are shown in Fig.\ref{fig:fig6}. We observe from the plots that the potentials corresponding to the models nearly overlap with each other at a lower value of $\phi$ but differ appreciably at higher values of the inflation field. This feature is expected since the inclusion of the higher order terms in $R$ in the model, automatically allows the entry of higher-order terms of the inflation field $\phi$. These extra higher order exponential terms
    play their crucial role in the potential in the vicinity of the starting of the inflation, which is $\phi_N=5.5 M_{Pl}$ for this model. This small change in potential and the consequent changes in its slope, and curvature associated with it give rise to an adequate variation in the 'significant digit' of the statistical data we are interested in. It is natural that these variations inevitably would have some influence in the process and that would certainly affect all the observables of interest, which have been established from our computation too. For an illustration, we are reporting only the calculations of scalar spectral index $n_s$ for $R2$, $R3$ and $R4$ models. We calculate the value of $n_s$ using Eqs.(\ref{ns},\ref{EPSILON},\ref{ETA},\ref{VR2},\ref{VR3}) and for the sake of comparison, we plot the values of $n_s$ for the three models. The plots are shown in Fig.\ref{fig:fig7}. In particular, if we calculate the value of $n_s$ at $\phi_N$ these come out as
    \[\begin{array}{ccc}
        {n_{s(R2)}} = 0.9690 = 0.9649 + 0.0041,\\
        {n_{s(R3)}} = 0.9683 = 0.9649 + 0.0034,\\
        {n_{s(R4)}} = 0.9662 = 0.9649 + 0.0013.
    \end{array}\]

    It is evident from the above data that although the results calculated for the models involving $R2$, $R3$, and $R4$ are consistent with the experimentally observed value of $n_s$(PLANCK)=$0.9649\pm 0.0042$, the addition of $R^3$, and $R^4$ terms in the model provide a better fit to the result. These variations in data can be found in all the inflationary model parameters; furthermore, improvements are also expected in presence of additional higher terms in the model. It is true that the addition of a higher-order term helps to give a better fit because of the added free parameter involved in it. But this process can not be extended at our will, because the number of free parameters will go on increasing along with the increasing difficulties of solving the higher-order equations. So one has to be extremely judicious during handling the higher-order the term, however, according to the desired accuracy it can be apprehended up to which order term is needed to include. But these considerations are beyond the scope of the present work reported in this paper. Some of the references of such calculations are given in the introduction of this article.
    \\

    \section{Summary and Discussions}
    In this work, we have attempted to construct an inflationary mode that
    contains both $R^3$ and $R^4$ terms. Both the terms are treated as
    perturbation over $R^2$ term which was introduced by Starobinsky in
    his seminal work. We have computed the exact expression of potential,
    and the slow-rolling parameters without any approximation. Note that
    in the article \cite{cheong2020beyond, huang2014polynomial},
    the calculations were carried out with the leading order term
    of the potential. Our endeavor in our work leads us to a successful realization that
    Starobinsky type slow-roll inflation is feasible even in the
    simultaneous presence of both the $R^3$ and $R^4$ terms when
    the exact form of the potential is used for computation. In
    order to achieve slow-roll along with a good agreement with
    the experimentally observed value $\Delta _R^2 = 2.1 \times {10^{ -9}}$
    the coefficient $\beta$, $\gamma$ and $\delta$ hade been found to be
    constrained. With the constants
    $\beta=10^{10}$ $\frac{\gamma}{\beta^2} = 10^{-5}$, $\frac{\delta}{\beta^3} =10^{-7}$
    we find that $r$ and $n_s$ comes out as $0.00242$ and $0.966242$
    respectively when $\phi_N =5.5M_{Pl}$ is set. These values are in
    agreement with recent Planck data. However, the simultaneous
    presence of both $R^3$ and $R^4$ makes the number of e-fold a
    little higher, $N=63$, than that was predicted in the original
    Starobinsky model, but it is very close to the admissible range
    of the value of the e-fold number. A systematic evaluation of
    reheating temperature, which involves the calculation of the the
    decay rate of inflation to the Standard Model particles has also
    been carried out to predict the minimum number of e-folds r
    equired\cite{liddle2000cosmological, liddle2003long, *martin2010first}
    in our model to solve the horizon problem. Our investigation establishes
    firmly that the inflation potential of this model does not destroy the
    classic characteristics of the Starobinsky model even after the inclusion
    of the term $R^3$ and $R^4$ in the original $R^2$-gravity model. All the
    interesting features of the  Starobinsky model is found to occur
    significantly in the presence of both the $R^3$ and $R^4$. The
    reason for this behavior lies in the fact that making the the
    amplitude of scalar power, $\Delta _R^2$,  'COBE normalized'
    suppresses the coefficients of  $R^3$ and $R^4$ in comparison
    to $R^2$ term. We also observe that the effect of the presence
    of $R^4$ is small indeed like the contribution of $R^3$ but its
    effect would be treated in the same footing with the contribution of $R^3$.

    We would like to reemphasize the summary of this work with some
    comments that are in order. $f(R)$ gravity theory with
    $R+R^2+R^3+R^4$ terms act as a consistent model of cosmological
    inflation. Reasonably satisfying agreement with experimental
    observations with the predictions of this model make it a
    phenomenologically viable one.
    \begin{acknowledgments}
        SA would like to thank Sourov Roy and Soumitra SenGupta of IACS, Kolkata
        for some valuable comments and suggestions. AR would like to acknowledge
        the facilities extended to him during his visit to the IUCAA, Pune.

    \end{acknowledgments}

    \bibliography{inflation21}

\end{document}